# Ultimate Limit of Biaxial Tensile Strain and N-Type Doping for Realizing an Efficient Low-Threshold Ge Laser

David S. Sukhdeo,[1] Shashank Gupta,[1] Krishna C. Saraswat,[1] Birendra (Raj) Dutt,[2,3] and Donguk Nam[*,4]

[1]Department of Electrical Engineering, Stanford University, Stanford, CA 94305, USA

[2]APIC Corporation, Culver City, CA 90230, USA

[3]PhotonIC Corporation, Culver City, CA 90230, USA

[4]Department of Electronics Engineering, Inha University, Incheon 402-751, South Korea

*dwnam@inha.ac.kr

## Abstract

We theoretically investigate how the threshold of a Ge-on-Si laser can be minimized and how the slope efficiency can be maximized in presence of both biaxial tensile strain and n-type doping. Our finding shows that there exist ultimate limits beyond which point no further benefit can be realized through increased tensile strain or n-type doping. Here were quantify these limits, showing that the optimal design for minimizing threshold involves about 3.7% biaxial tensile strain and $2 \times 10^{18}$ cm$^{-3}$ n-type doping, whereas the optimal design for maximum slope efficiency involves about 2.3% biaxial tensile strain with $1 \times 10^{19}$ cm$^{-3}$ n-type doping. Increasing the strain and/or doping beyond these limits will degrade the threshold or slope efficiency, respectively.

**Introduction**

While optical interconnects offer compelling performance advantages over existing copper interconnects [1], manufacturing an optical link in a way that is compatible with existing silicon (Si) CMOS technology is a serious challenge [2]. Si itself is inherently unsuitable for light emission due to its complete absence of a direct bandgap [3] and there are many manufacturing challenges associated with integrating III-V materials with standard Si CMOS electronics, though progress is being made on the latter issue [4]–[6]. In order to bypass these limitations researchers have sought to use Group IV materials for optoelectronic applications as these materials pose far fewer contamination concerns when integrated with existing Si CMOS electronics [7], [8]. In this framework germanium (Ge) has long found use as a detector [9]–[12] and more recently as a modulator [13], [14]. The use of Ge is highly advantageous in this context since Ge can be readily grown on Si through high-quality heteroepitaxy [15], [16] and because Ge is already widely used in commercial CMOS processing [17], [18]. Moreover, although Ge has an indirect bandgap of 0.667 eV, it also has a direct bandgap of 0.8eV [19] which can be readily accessed for absorptive applications such as detectors [9]–[12] and modulators [13]. For light emission applications, however, the indirect bandgap poses a considerably larger hurdle: over 99.98% of injected electron will reside in the indirect conduction valleys [20] where they cannot contribute substantially to useful optical processes [21], [22]. This makes building an efficient Ge light emitter, and in particular a Ge laser, quite challenging [8], [23].

Interest in band-engineered Ge for light emission took off in 2007 with a theoretical publication [21] which claimed that a combination of a small 0.25% biaxial tensile strain and 7.6x10$^{19}$ cm$^{-3}$ n-type doping can achieve up to 400 cm$^{-1}$ optical gain in an ideal structure, enough to build a working laser [21]. Soon afterward, researchers first began looking into the new possibility of further increasing the tensile strain for reducing the threshold of a Ge laser because tensile strain can reduce the energy difference between the direct Gamma and indirect L valleys. In 2012, researchers presented in-depth theoretical modeling where they compared the two techniques, n-type doping and tensile strain engineering, to determine which one of them is more suitable for realizing an efficient low threshold on-chip Ge laser. It was found that without tensile strain, the performance improvement of a Ge laser can be significantly limited and that it would be highly desirable to use both biaxial tensile strain and n-type doping. Ever since then, much experimental progress has already been made for n-type doping and tensile strain. For example, n-type doping of ~$1 \times 10^{20}$ cm$^{-3}$ has been demonstrated in [24] and, separately, biaxial tensile strain of ~1.5% has been demonstrated [25]. While experimental work on band-engineered Ge continues, it is not yet studied if there are any ultimate limits beyond which tensile strain and/or n-type doping would not improve the performance of a Ge laser. Therefore, in this paper, we theoretically investigate the ultimate limits of tensile strain and n-type doping and suggest a roadmap towards the ultimate performance improvements of a Ge laser. We find that tensile strain remains useful for threshold reduction up until ~3.7% biaxial strain, which is nearly enough to make Ge a zero-bandgap material, and the best n-type doping at this value is ~$2 \times 10^{18}$ cm$^{-3}$. However, we find that the ultimate limit for improving slope efficiency occurs much sooner at ~2.3% biaxial strain and ~$1 \times 10^{19}$ cm$^{-3}$ n-type doping, and pushing the strain higher than ~2.3% results in very poor slope efficiency. This result is explained by the increased free carrier absorption at the redshifted emission wavelengths associated with very highly strained Ge and is a fundamental limitation which arises even in our ideal parasitic-free model. On a practical level this means that increasing the tensile strain from present values [25]–[27] would be very helpful, but slope efficiency will eventually become a more fundamental limitation than the high thresholds [28]–[30] which currently plague band-engineered Ge lasers.

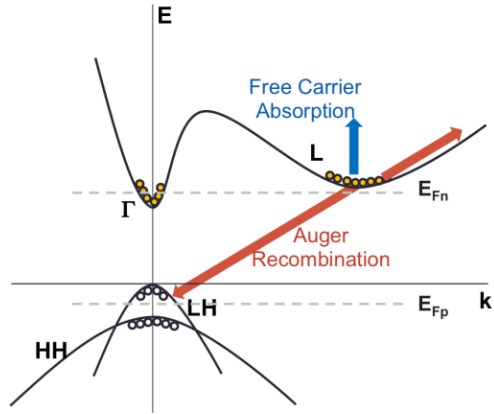

*Fig. 1.* *Illustration of the free carrier absorption and Auger recombination processes in Ge.*

Before delving into quantitative modeling, we first explain the possible carrier actions that happen in highly strained and heavily n-type doped Ge to explain how the threshold can be changed in the presence of strain and n-type doping. Fig. 1 illustrates the band structure of strained Ge with >2.4% biaxial tensile strain and thus a direct bandgap. Due to n-type doping, we expect a large electron concentration in the indirect L conduction valley. Quasi Fermi levels that are within the conduction and valence bands, as illustrated in Fig. 1, indicate population inversion in Ge, however lasing cannot occur until the gain from the direct transition exceeds the loss from free carrier absorption. This not only raises the threshold due to increased pumping needed to overcome these losses, but also represents a very substantial drain on slope efficiency as many generated photons will be lost to this free carrier absorption process before they can exit the cavity as useful light emission.

We can now begin quantifying how tensile strain affects the performance of a Ge laser by considering how strain affects Ge's band structure. First, when tensile strain is increased, the energy difference between the direct $\Gamma$ and indirect L valleys become smaller, and eventually the band gap will become direct. Because a larger fraction of injected electrons will populate the lower direct $\Gamma$ valley with larger tensile strain, one can expect that the optical gain the direct transition to increase with tensile strain for a constant carrier injection level. Also, because strain reduces the bandgap energy, the lasing wavelength will be redshifted as modeled in Ref [31], [32]. Interestingly, at these longer emission wavelengths, the free carrier loss also increases because free carrier losses have a strong wavelength dependence according to the empirical fit of Ref. [21] which we show here as Equation (1):

$$\alpha_{FCA} = 3.4 \times 10^{-25} n\lambda^{2.25} + 3.2 \times 10^{-25} p\lambda^{2.43} \quad (1)$$

For the optical net gain these two components, gain from the direct transition and free carrier loss, compete against each other to determine the optical net gain, which is just the gain from the direct transition minus the free carrier loss. Therefore, as the tensile strain in Ge is increased, the gain from the direct transition increases due to increased occupancy of the gamma conduction valley but the free carrier loss also increases due to the redshifted emission wavelengths. It is important to carefully examine which of these two effects term dominates for the positive optical net gain, and we will show in this work that there is an ultimate limit for tensile strain beyond which too much strain will actually be very harmful. A similar story applies when the n-type doping is increased. While introducing moderate n-type doping is helpful due to increased occupancy of the gamma conduction valley, too much doping can be harmful because eventually all the relevant states in the indirect valley have been filled and so adding more extrinsic electrons serves only to worsen the free carrier absorption [31] . Thus, we find that there is an ultimate limit beyond which too much n-type doping will be quite harmful. In this work we will explore the ultimate limits of biaxial tensile strain and n-type doping with respect to both threshold and slope efficiency, and also carefully examine the interaction between strain and doping with respect to these ultimate limits.

To quantitatively investigate the ultimate limit of strain and doping, we first calculate the bandstructure of biaxially strained Ge. The nearest neighbor $sp^3d^5s*$ empirical tight-binding model is used in this work to obtain the bandstructure in the first Brillouin zone, and the use of tight-binding ensures that the valence band mixing and warping under strain is fully accounted for in our model. To examine the ultimate limit, we perform detailed modeling for biaxial strains of up to 4% which is much farther than any previous work [31], [33]–[37]. Fig. 2 shows how band edges change with strain. At 4.0% strain the bandgap shrinks to under 0.02eV and at 4.1% strain Ge becomes a negative bandgap material according to our tight-binding model.

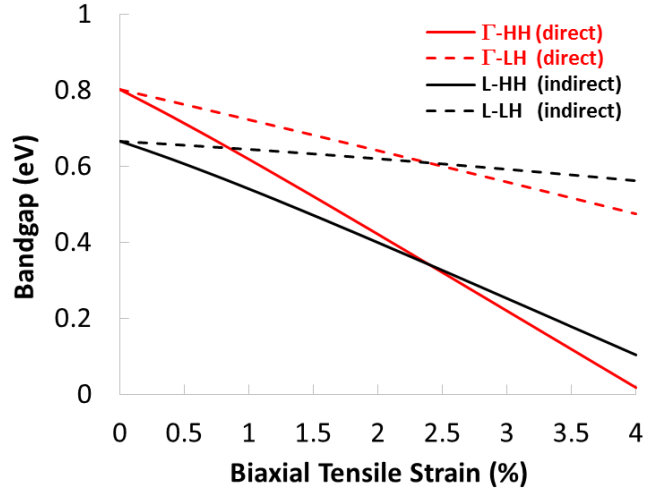

*Fig. 2.* Ge's direct (Γ) and indirect (L) bandgap energies vs. biaxial tensile strain according to our tight-binding model. Crossover of the direct gap is visible at 2.4% biaxial strain.

Based on the calculated bandstructure, we perform laser modeling following the same approach as in Ref. [31]. We assume an empirical absorption coefficient for Ge that accounts for valence band splitting given by Equation 2 [38]:

$$\alpha_\Gamma = 1.9 \times 10^4 \text{ eV}^{0.5}\text{cm}^{-1} \left( 0.682\sqrt{E_{photon} - E_g^{\Gamma-HH}} + 0.318\sqrt{E_{photon} - E_g^{\Gamma-LH}} \right) \Big/ E_{photon} \quad (2)$$

The above equation divides the total absorption into its two major components, one for each of the Γ-LH and Γ-HH transitions, in the ratio of their joint density of states. The optical gain from these direct transitions is then computed by multiplying the absorption coefficient with the Fermi inversion factor as shown in Equation (3):

$$\gamma_\Gamma = \alpha_\Gamma(f_c - f_v) \qquad (3)$$

where $f_c$ and $f_v$ are electron and hole quasi Fermi levels respectively. The optical net gain is then obtained by subtracting the free carrier absorption given by Equation (1) from the optical gain in Equation (3). The quasi Fermi levels used in Equation (3) allow us to compute the corresponding injected carrier density according to the Fermi-Dirac statistics. The injected carrier density is then converted to a drive current density using the continuity equation. The coefficients for direct and Auger recombination are taken from Ref [21]. Threshold current density is then simply the drive current needed to achieve an optical net gain equal to the presume cavity loss, and the computed threshold current density is shown in Fig. 3 as a function of biaxial tensile strain and n-type doping. As shown in Fig. 3, our model predicts that the ideal combination of strain and doping for minimum threshold is ~3.7% biaxial tensile strain with ~$2\times10^{18}$ cm$^{-3}$ n-type doping. At this combination of 3.7% strain and $2\times10^{18}$ cm$^{-3}$ n-type doping, we expect a threshold of only ~80 A/cm$^2$, a reduction of >5000x compared to the ~600 kA/cm$^2$ threshold that our model predicts for state-of-the-art Ge laser values of 0.2% and $5\times10^{19}$ cm$^{-3}$ n-type doping.

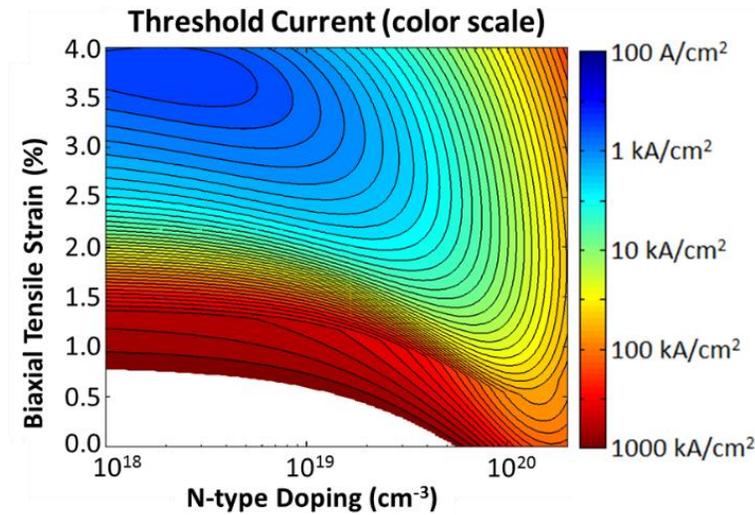

*Fig. 3. Threshold current density vs. biaxial tensile strain and n-type doping at 700 cm$^{-1}$ optical cavity loss, assuming a double heterostructure with a 300nm thick Ge active region.*

To investigate the ultimate limit of n-type doping for reducing the threshold of a Ge laser, we re-plot Fig. 3 as a 2D plot for representing the threshold vs. n-type doping for various strain values as shown in Fig. 4(a). It is clearly shown therein that there exists ultimate limits of n-type doping for each strain value beyond which point n-type doping start increasing the threshold. This is because the increased free carrier loss from heavier n-type doping eventually starts to exceed the gain from the direct transition, which results in reduced optical net gain with the same current injection level. This ultimate limit of n-type doping becomes lower with larger tensile strain because the benefit of n-type doping on the optical gain is smaller while the free carrier loss becomes more rapidly increased with doping due to increased emission wavelength. This phenomenon of the ultimate limit for doping decreasing in the presence of strain is also shown explicitly in Fig. 4(b).

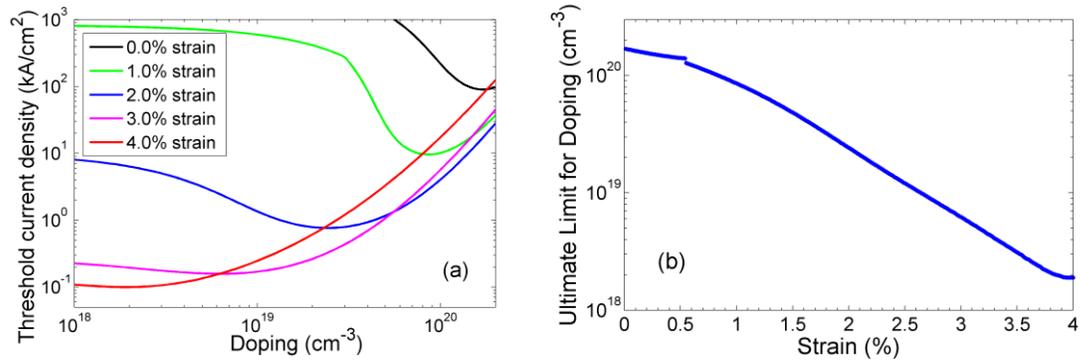

*Fig. 4. (a) Threshold current density of a 300nm-thick double heterostructure Ge laser vs. n-type doping for different amounts of biaxial strain. (b) Ultimate limit for doping vs. biaxial tensile strain. In all cases the optical cavity loss is assumed to be 700 cm$^{-1}$ with a defect-assisted minority carrier lifetime of 100 ns.*

We also look into the ultimate limit of biaxial tensile strain for a low-threshold Ge laser. Fig. 5 shows the threshold vs. strain for various n-type doping values. It is also found that there exist ultimate limits of biaxial tensile strain for each doping level beyond which there is no performance enhancement from further tensile strain. This ultimate limit comes from the fact that increased free carrier loss due to increased emission wavelength dominates the improved optical gain from strain. Similar for the case of n-type doping, the ultimate limit values for strain become smaller with higher n-type doping. Fig. 5(b) shows how the ultimate limit for strain changes with n-type doping.

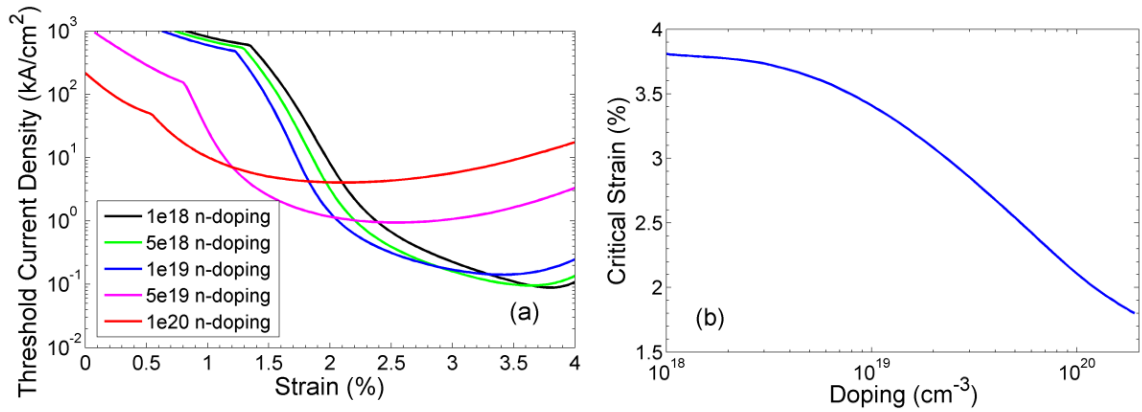

*Fig. 5. (a) Threshold current density of a 300nm-thick double heterostructure Ge laser vs. biaxial tensile strain for different amounts of n-type doping. (b) Ultimate limit for strain vs. biaxial tensile strain. In all cases the optical cavity loss is assumed to be 700 $cm^{-1}$ with a defect-assisted minority carrier lifetime of 100 ns.*

While the threshold is one particularly critical figure of merit for realizing a practical Ge laser, another key parameter is the slope efficiency which is defined as the differential power efficiency of the proposed laser just above threshold [39]. Following similar approach as in Ref. [31], we calculate the slope efficiency to find if there exist ultimate limits for strain and doping. For this calculation, we presume the optical cavity loss to be 700 $cm^{-1}$. As shown in Fig. 6, the maximum slope efficiency occurs at ~2.3% strain with ~$1\times10^{19}$ $cm^{-3}$ n-type doping. This means that for ~$1\times10^{19}$ $cm^{-3}$ n-type doping, it is not desirable to pursue strain greater than 2.3% if slope efficiency is important. While the minimum threshold occurs at ~3.7% it would be unwise to actually implement such a strain as it would result in a slope efficiency of only about 15% even before considering parasitics such as contact resistance, surface scattering, optical losses in the electrodes, etc. At 2.3% strain and $1\times10^{19}$ cm$^{-3}$ doping combination, the slope efficiency is ~47% before parasitics with a threshold of only ~500 A/cm$^3$. While this threshold is about 6x higher than what might be achieved at 3.7% strain (at the expense of very poor slope efficiency), it still represents a >1000x threshold reduction compared to the present state-of-the-art Ge laser parameters of 0.2% biaxial strain and $5\times10^{19}$ cm$^{-3}$ n-type doping.

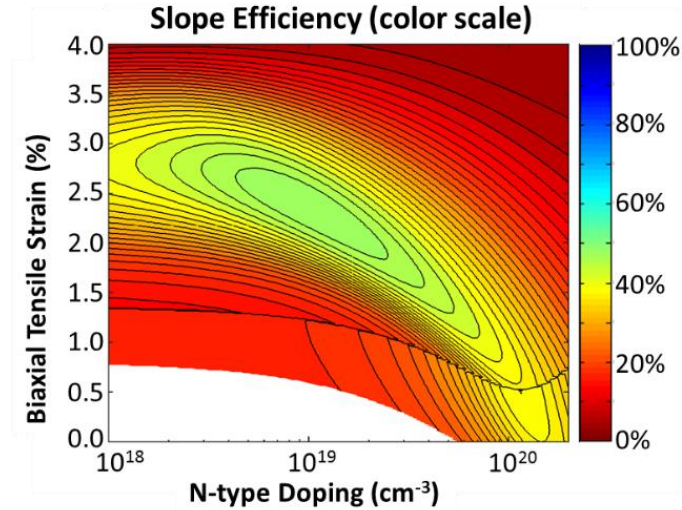

***Fig. 6.*** *Slope efficiency vs. biaxial tensile strain and n-type doping at 700 cm$^{-1}$ optical cavity loss.*

It is also worthwhile to consider the mechanism by which the slope efficiency is constrained to this ultimate limit. The slope efficiency consists of two components that combine multiplicatively: the differential quantum efficiency and the photon energy to electron energy ratio as shown in Equation (4):

$$\text{Slope efficiency} = \text{differential quantum efficiency} \times \left(E_{\text{photon}}/E_{\text{electron}}\right)$$

$$= \frac{\alpha_{\text{cavity}}}{\alpha_{\text{cavity}} + \alpha_{\text{FCA}}} \times \frac{hc/\lambda}{E_{Fn} - E_{Fp}} \qquad (4)$$

where $\alpha_{cavity}$ is the optical cavity loss, $\alpha_{FCA}$ is the free carrier absorption, h is the Planck constant, c is the speed of light, $\lambda$ is the lasing wavelength , and $E_{Fn} - E_{Fp}$ is quasi-Fermi level separation. By computing each of the two components separately, we find that slope efficiency is almost exclusively limited by the differential quantum efficiency rather than by the photon-to-electron energy ratio as shown in Fig. 7. In fact, except for a small region in the upper right portion of Fig. 7(b), i.e. when the doping is anyway much larger than its optimal value, the photon energy to electron ratio is consistently about 85-90% across all strain and doping values. The differential quantum efficiency (Fig. 7(a)) on the other hand shows a strong dependence on the strain and doping, with a 52% differential quantum efficiency at ~2.3% strain and ~$1\times10^{19}$ cm$^{-3}$ n-type doping that decreases rapidly if the strain is further increased. Since the differential quantum efficiency is the limiting factor of the slope efficiency, it logically follows that this ultimate limit for slope efficiency is governed by free carrier absorption which increases with strain due to the redshifted emission wavelength in accordance with Equation (1).

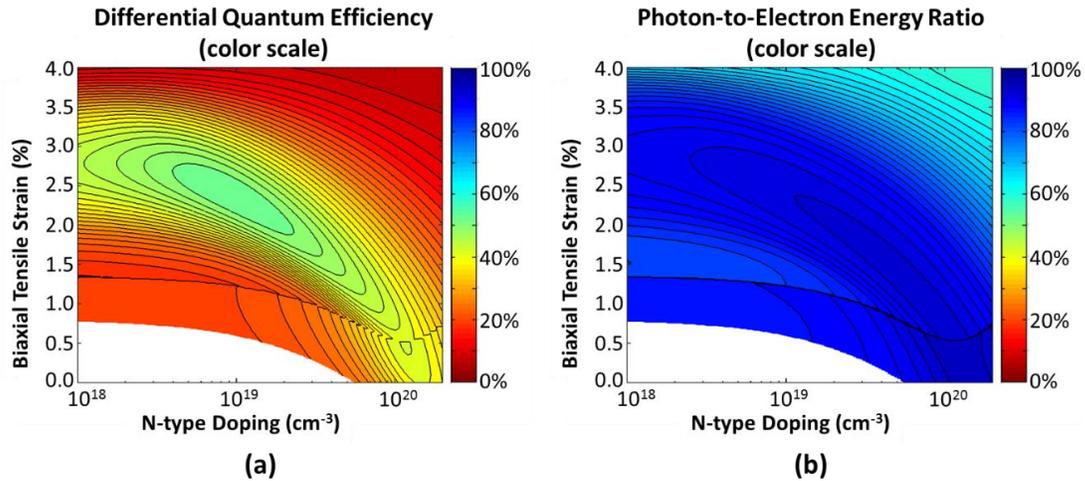

*Fig. 7. Components of the slope efficiency, computed at 700 cm$^{-1}$ optical cavity loss. (a) Differential quantum efficiency, (b) ratio of the output photon energy to the input electron energy. In all cases a double heterostructure design is assumed.*

## Conclusion

We have investigated the ultimate limits of biaxial tensile strain and n-type doping for improving the performance of a Ge laser. Using tight-binding modeling and numerical calculations, we compute the threshold and slope efficiency and carefully examine how the two figures of merit are affected by both strain and doping. We find that there clearly exist the ultimate limits for both strain and doping beyond which point there is no benefit from pursuing further strain and/or doping. We attribute this presence of the ultimate limits to the competition between the increased optical gain and the increased free carrier loss due to strain and doping. Most interestingly, we find that the ultimate limit is very different for threshold than it is for slope efficiency. For threshold optimization the ultimate limit occurs at 3.7% biaxial tensile strain and $1\times10^{18}$ cm$^{-3}$ n-type doping, which would result in an ~80 A/cm$^2$ threshold and a 15% slope efficiency. For slope efficiency, on the other hand, the ultimate limit occurs at only ~2.3% biaxial tensile strain and ~$1\times10^{19}$ cm$^{-3}$ n-type doping, resulting in a higher threshold of ~500 A/cm$^2$ but also a dramatically higher slope efficiency of ~47%. The key takeaway from these results is that, when slope efficiency is an important figure of merit, there is not much practical incentive to push the strain beyond ~2.3%, and doing so would could actually result in a very inefficient laser. For applications where a minimum threshold is of utmost importance, however, higher strains may be warranted. Through these results, we present guidance for researchers how much strain and doping should be pursued for threshold minimization or slope efficiency maximizing depending on the particular application intended.

## Acknowledgements

This work was supported by the Office of Naval Research (grant N00421-03-9-0002) through APIC Corporation (Dr. Raj Dutt) and by a Stanford Graduate Fellowship. This work was also supported by an INHA UNIVERSITY Research Grant and by the Pioneer Research Center Program through the National Research Foundation of Korea funded by the Ministry of Science,

ICT & Future Planning (2014M3C1A3052580). The authors thank Boris M. Vulovic of APIC Corporation for helpful discussions. The authors also thank Ze Yuan of Stanford University for his help implementing the tight-binding code.